\documentclass[aps,prd,twocolumn,showpacs,nofootinbib]{revtex4-1}

\usepackage{amssymb} \usepackage{amsmath} \usepackage{graphicx}
\usepackage{epsfig,latexsym}

\RequirePackage{xspace} \allowdisplaybreaks

\begin{document}

\def\bef{\begin{figure}}
\def\eef{\end{figure}}

\newcommand{\nl}{\nonumber\\}

\newcommand{\ans}{ansatz }
\newcommand{\be}[1]{\begin{equation}\label{#1}}
\newcommand{\beq}{\begin{equation}}
\newcommand{\ee}{\end{equation}}
\newcommand{\beqn}[1]{\begin{eqnarray}\label{#1}}
\newcommand{\eeqn}{\end{eqnarray}}
\newcommand{\bd}{\begin{displaymath}}
\newcommand{\ed}{\end{displaymath}}
\newcommand{\mat}[4]{\left(\begin{array}{cc}{#1}&{#2}\\{#3}&{#4}
\end{array}\right)}
\newcommand{\matr}[9]{\left(\begin{array}{ccc}{#1}&{#2}&{#3}\\
{#4}&{#5}&{#6}\\{#7}&{#8}&{#9}\end{array}\right)}
\newcommand{\matrr}[6]{\left(\begin{array}{cc}{#1}&{#2}\\
{#3}&{#4}\\{#5}&{#6}\end{array}\right)}
\newcommand{\cvb}[3]{#1^{#2}_{#3}}
\def\lsim{\raise0.3ex\hbox{$\;<$\kern-0.75em\raise-1.1ex
e\hbox{$\sim\;$}}}
\def\gsim{\raise0.3ex\hbox{$\;>$\kern-0.75em\raise-1.1ex
\hbox{$\sim\;$}}}
\def\abs#1{\left| #1\right|}
\def\simlt{\mathrel{\lower2.5pt\vbox{\lineskip=0pt\baselineskip=0pt
           \hbox{$<$}\hbox{$\sim$}}}}
\def\simgt{\mathrel{\lower2.5pt\vbox{\lineskip=0pt\baselineskip=0pt
           \hbox{$>$}\hbox{$\sim$}}}}
\def\unity{{\hbox{1\kern-.8mm l}}}
\newcommand{\eps}{\varepsilon}
\def\ep{\epsilon}
\def\ga{\gamma}
\def\Ga{\Gamma}
\def\om{\omega}
\def\omp{{\omega^\prime}}
\def\Om{\Omega}
\def\la{\lambda}
\def\La{\Lambda}
\def\al{\alpha}
\newcommand{\ov}{\overline}
\renewcommand{\to}{\rightarrow}
\renewcommand{\vec}[1]{\mathbf{#1}}
\newcommand{\vect}[1]{\mbox{\boldmath$#1$}}
\def\tm{{\widetilde{m}}}
\def\mcirc{{\stackrel{o}{m}}}
\newcommand{\Dm}{\Delta m}
\newcommand{\dm}{\varepsilon}
\newcommand{\tanb}{\tan\beta}
\newcommand{\nbar}{\tilde{n}}
\newcommand\PM[1]{\begin{pmatrix}#1\end{pmatrix}}
\newcommand{\up}{\uparrow}
\newcommand{\down}{\downarrow}
\def\omE{\omega_{\rm Ter}}
%

\newcommand{\Dsusy}{{susy \hspace{-9.4pt} \slash}\;}
\newcommand{\DCP}{{CP \hspace{-7.4pt} \slash}\;}
\newcommand{\mc}{\mathcal}
\newcommand{\gr}{\mathbf}
\renewcommand{\to}{\rightarrow}
\newcommand{\gtc}{\mathfrak}
\newcommand{\wh}{\widehat}
\newcommand{\br}{\langle}
\newcommand{\kt}{\rangle}


\def\lsim{\mathrel{\mathop  {\hbox{\lower0.5ex\hbox{$\sim$}
\kern-0.8em\lower-0.7ex\hbox{$<$}}}}}
\def\gsim{\mathrel{\mathop  {\hbox{\lower0.5ex\hbox{$\sim$}
\kern-0.8em\lower-0.7ex\hbox{$>$}}}}}

\def\nn{\\  \nonumber}
\def\de{\partial}
\def\brf{{\mathbf f}}
\def\bbf{\bar{\bf f}}
\def\bF{{\bf F}}
\def\bbF{\bar{\bf F}}
\def\bA{{\mathbf A}}
\def\bB{{\mathbf B}}
\def\bG{{\mathbf G}}
\def\bI{{\mathbf I}}
\def\bM{{\mathbf M}}
\def\bY{{\mathbf Y}}
\def\bX{{\mathbf X}}
\def\bS{{\mathbf S}}
\def\bb{{\mathbf b}}
\def\bh{{\mathbf h}}
\def\bg{{\mathbf g}}
\def\bla{{\mathbf \la}}
\def\bmu{\mathbf m }
\def\by{{\mathbf y}}
\def\bmu{\mbox{\boldmath $\mu$} }
\def\bsig{\mbox{\boldmath $\sigma$} }
\def\bunity{{\mathbf 1}}
\def\cA{{\cal A}}
\def\cB{{\cal B}}
\def\cC{{\cal C}}
\def\cD{{\cal D}}
\def\cF{{\cal F}}
\def\cG{{\cal G}}
\def\cH{{\cal H}}
\def\cI{{\cal I}}
\def\cL{{\cal L}}
\def\cN{{\cal N}}
\def\cM{{\cal M}}
\def\cO{{\cal O}}
\def\cR{{\cal R}}
\def\cS{{\cal S}}
\def\cT{{\cal T}}
\def\eV{{\rm eV}}
%

\title{Aspects of quantum chaos inside black holes}

\author{Andrea Addazi$^1$}\email{andrea.addazi@infn.lngs.it}
\affiliation{$^1$ Dipartimento di Fisica,
 Universit\`a di L'Aquila, 67010 Coppito AQ and
LNGS, Laboratori Nazionali del Gran Sasso, 67010 Assergi AQ, Italy}

\begin{abstract}

We will argument how infalling information can be chaotized inside realistic quantum black holes.

\end{abstract}

\maketitle

\section{Introduction and Conclusions}

Theoretical physicists are all agreed that
Semiclassical Black holes are paradoxical objects 
(as nicely reconfirmed by several discussions during the Karl Schwarzschild meeting 2015).
However, a clear strategy in order to solve this problem is still unknown. 

In this paper, we would like to suggest that 
infalling information could be chaotized inside a black hole .
Our claim is related to a different picture about 
quantum black holes' nature: 
 we retained unmotivated to think seriously about a quantum black hole
as a conformal Penrose's diagram, {\it i.e} as a smoothed semiclassical geometry 
with a singularity in its center (eventually cutoff at the planck scale). 
In particular, one could expect that, in a "window" of lenght scales 
among the Schwarzschild radius and the Planck scale, 
there is 
a non-topologically trivial region of space-time rather than a smoothed one.
A realistic black hole 
could be a superposition of 
different horizonless solutions, 
perhaps associated GR gravitational instantons or "exotic" gravitational instantons
\footnote{In string theory, the class of instantons is much larger than in field theories. 
Applications of a particular class of these solutions in particle physics were recently studied in 
\cite{Addazi:2015gna,Addazi:2014ila,Addazi:2015ata,Addazi:2015rwa,Addazi:2015hka,Addazi:2015fua,Addazi:2015oba,Addazi:2015goa,Addazi:2015yna,Addazi:2015ewa}
.}.
In this picture, a black hole' horizon is an approximated Chauchy null-like surface
(for energy scales closed to an inverse Schwarzschild radius).
However, for lenght scales $L$ in the range $l_{Pl}<<L<<R$,
geometrical deviations and asperities with respect to semiclassical smoothed geometries 
are reasonable expected. 
In this regime, gravitational interactions among horizonless geometries 
can be neglected as well as microscopical exchanges of matter and gauge fields 
among their surfaces.  
In this sense, a black hole cannot be described by a single Penrose's diagram 
at all the lenght scales. In particular, in the "middle region" 
a black hole would be described by a superposition of a
large number of Penrose's diagrams. 

Such a black hole can be rigorously defined in an euclidean path integral formulation. 
It emits a thermal radiation like a semiclassical 
BH, with small corrections on Bekenstein-Hawking entropy (see section II) \cite{Addazi:2015gna}. 

At this point, a further  question is the following:
what happen to infalling informations in such a "scale variant" system?
Let us consider the usual thought experiment of a infalling radiation 
in a quantum pure state, with a very small initial frequency $\omega \simeq R^{-1}$. 
Such a radiation will start to probe a smoothed semiclassical geometry of a black hole, 
near the horizon. 
However, radiation will be inevitably blueshifted inside
the gravitational potential of a black hole,
{\it i.e} its De Broglie wave length starts to be smaller 
than the Schwarzschild's radius. 
So that, infalling radiation will start to probe the middle region
before than the full quantum quantum regime. 
In the middle region, radiation is scattered back and forth among asperities
that usually are not present at all in semiclassical BH solutions. 
As a consequence, radiation will be chaotically diffracted inside 
this system. At that middle scales, a black hole is a sort 
of {\it space-temporal chaotic Sinai billiard} rather than a smoothed manifold. 
Usually, in simpler classical chaotic billiards than our one, 
chaotic zones of unstable orbits trapped forever in the system are formed.
Simple examples of such a trapped paths: 
i) an orbit trapped in back and forth scatterings among the asperity A and the asperity B (AB segments);
ii) one trapped among $A,B,C$ asperities (triangular orbits);
and so on.  
Considering quantum fields rather than classical trajectories, 
one has also to consider quantum transitions 
induced by inelastic scatterings on gravitational backgrounds $<g,...,g>$ (thought as a vacuum expectation value of 
gravitons). 
$\phi +<g,...,g>\rightarrow X+<g,...,g>$
where $\phi$ is a generic gauge/matter field, 
and $X$ is a collection of $N$ fields. 
For example a process like a photon-background scattering
$$\gamma +<g,...,g> \rightarrow q\bar{q}+<g,...,g>\rightarrow hadronization +<g,...,g>$$
will lead to a complicated hadronic cascade of entangled fields. 
As a consequence, such a system is even more chaotic than 
classical one. 
So that, 
a part of the initial infalling information is effectively fractioned in
a "forever" (black hole lifetime or so) trapped part and another one, 
so that 
$$|IN\rangle =a|OUT\rangle +b|TRAPPED\rangle$$
where $a,b$ parametrize our ignorance about the 
space-time billiard, 
$|OUT \rangle$ is emitted as Bekestein-Hawking radiation. 
As a consequence, the in-going information 
is a linear combination of outgoing informations 
and trapped informations
during $0<<t<<t_{Evaporation}$. 
In this picture, information paradoxes are understood as 
an {\it apparent} losing of unitarity. 
In fact, $| IN \rangle \rightarrow |OUT \rangle$ 
 is not allowed by quantum mechanics:
 $|IN\rangle$ is a pure state, 
 while $|OUT \rangle$ is a mixed one. 
 However, also $|TRAPPED \rangle$ is a mixed state, 
 and a linear combination of two mixed states can be a pure one. 
 In this approach, a $|IN\rangle \rightarrow |OUT \rangle$ transition can be 
 effectively described in a density matrix approach, with an effective non-unitary 
 evolution. 
 However, unitarity is not lost at fundamental level because of 
 the real transition $|IN\rangle \rightarrow a|OUT \rangle+b|TRAPPED\rangle$
 is not contradicting unitarity. 
 Let us consider, for example, a (famous) Bekenstein-Hawking particle-antiparticle pair 
 created nearby the black hole horizon. 
 As usual, one the two is captured inside the black hole space-like interior, 
 while the second one can tunnel outside the horizon. 
As well known, the two particles are entangled, and this will lead to 
the undesired firewall paradox. 
 However, in a frizzy black hole, the infalling pair will start to be blueshifted 
 so that it will start to scatter back and forth inside the system, 
 giving rise to an exponentially growing cascade of N particles
 continuing to scatter and to scatter in the billiard. 
The process will be even more chaotic in a realistic case in which 
a large number of infalling partners from a large number of Bekenstein-Hawking pairs have to be considered. 
As a consequence, P outgoing pairs will be entangled with a total number $N>>P$ 
 of particles inside the system. This practically disentangles the $P$ outgoing pairs 
 from the original ones, as a quantum decoherence effect induced by the non-trivial space-time topology.  
 In other words, the space-time topology is collapsing the entangled wave function
 as a quantum decoherence phenomena, as well as two entangled pairs are disentangled by a an experimental 
 apparatus. 
 The entanglement entropy is linearly growing with the number of back and forth scatterings $n$
of a particle, because of the density matrix of the internal black states are exponentially growing with $n$:
$$S_{interior}=-Tr \rho_{interior}log S_{interior}\sim n$$
so that is growing with time. 
On the other hand, for $P$ Bekenstein-Hawking particles 
$S_{int}\sim n \,log\,P$.
Our model predicts $S_{B.H.}\sim P$ from entanglement entropy definition.

However, if a frizzy black hole emits a Bekenstein-Hawking radiation with small deviations from thermality, 
it cannot have an infinite life-time. On the other hand, the non-trivial topological space-time configuration 
of a frizzy black hole is sourced by the black hole mass. 
The final configuration after the complete black hole evaporation is a Minkowski space-time 
with a dilute residual radiation. 
As a consequence, a space-time phase transition from the "frizzy" topology to the Minkowski space-time 
is expected at the Page time or so. As a consequence, chaotic saddles of trapped information 
will be emitted in the environment as a {\it final information burst}.
For this motivation, the S-matrix describing BH evolution from the initial collapse/formation
to its complete evaporation is unitary: 
$$\langle COLLAPSE|S|EVAPORATION\rangle $$
$$=\langle TOTAL INFALLING| S| (a|TRAPPED\rangle +b |OUT\rangle)$$
The trapped probability density  $\rho(T)$ is approximately described by
$$\frac{d\rho(T)}{dT}\sim -\frac{1}{T^{2}}e^{-\Gamma T }$$
In fact, $\rho(T)$ 
is dependent by
the number of asperities $N_{s}$ as $\rho \sim N_{s}e^{-\Gamma T}$,
where $\Gamma$ is proportional to {\it effective average deepness} of 
 asperities (trapping $\rho$). But the number of asperities is depending by the Black hole mass. 
In turn, the black hole mass  decreases with the temperature as $dM/dT=-1/8\pi T^{2}$. 

To conclude, chaotic aspects of quantum black holes could be relevantly connected to 
the information paradoxes. In particular, a semiclassical black hole could be reinterpreted 
as a superposition of horizonless geometries, chaotizing infalling informations.
Such an approach could have surprising connections 
with recent results in contest of AdS/CFT correspondence \cite{C0}.

 \section{Euclidean path integral of a frizzy black hole}
 
{\bf Definition}:
let us consider a generic system of  N horizonless background metrics (suppose to be eliminated at the Planck scale)
inside a box with a surface $\partial \mathcal{M}$.
 This system is defined {\it frizzy black hole} if it satisfies the following hypothesis:
 
 i) A formal definition of partition functions $Z_{I}$ for each metric tensor $g^{I=1,...,N}$
cab be defined. The N metrics are in thermal equilibrium with the box.

 ii) In semiclassical regime, the leading order of the total partition 
function associated to this system is the product of the single partition function:
$$Z_{TOT}= \prod_{I=1}^{N} Z_{I}$$
\footnote{
This approximation can be trusted if and only of 
intergeometries' interactions are small  
with respect to the temperature inside the box.}

iii) The total average partition function has a form
$$\langle Z_{TOT}\rangle=e^{-\frac{\beta^{2}}{16\pi}-\frac{\sigma_{\beta}^{2}}{16\pi}}=Z_{E}e^{\frac{-\sigma_{\beta}^{2}}{16\pi}}$$
where $Z_{E}$ is the usual semiclassical euclidean partition function, 
$\sigma_{\beta}$ the variance of $\beta$-variable in the system.
This corresponds to an entropy 
$$\langle S \rangle=\frac{\beta^{2}}{16\pi}+\frac{\sigma_{\beta}^{2}}{16\pi}$$

  \section{Semiclassical chaotic scattering on a Space-time Sinai Biliard}

An effective non-relativistic quantum mechanical approach 
is not fully valid and motivated. 
However, we will show some general proprieties suggested by this approach:
these can be extended to the more realistic problem. 

Let us remind, that $\psi_{t}(\vect{r})$ 
is obtained by an initial $\psi_{0}(\vect{r}_{0})$ 
by the unitary evolution 
$$\psi_{t}(\vect{r})=\int d\vect{r}_{0}K(\vect{r},\vect{r}_{0},t)\psi_{0}(\vect{r}_{0})$$
where $K$ is 
$$K(\vect{r},\vect{r}_{0},t)=\int \mathcal{D}\vect{r} e^{\frac{i}{\hbar}I}$$
and
$$I=\int_{0}^{t} dt L(\vect{r},\dot{\vect{r}})$$
and $L$ is the lagrangian of a particle. 
The semiclassical limit corresponds to 
$$I=\int_{0}^{t}[\vect{p}\cdot d\vect{r}-Hd\tau]>>\hbar$$
In this regime, 
 the leading contribution to the path integral 
is just given by classical chaotic trapped orbits. 
The semiclassical propagator
can be written as
$$K_{WKB}(\vect{r},\vect{r}_{0},t)\simeq \sum_{n}\mathcal{A}_{n}(\vect{r},\vect{r}_{0},t)e^{\frac{i}{\hbar}I_{n}}$$
where we are summing on all over the classical orbits of the system, 
while amplitudes $\mathcal{A}_{n}$
are 
$$\mathcal{A}_{n}(\vect{r},\vect{r}_{0},t)=\frac{1}{(2\pi i \hbar)^{\nu/2}}\sqrt{|det[\partial_{\vect{r}_{0}}\partial_{\vect{r}_{0}}I_{n}[\vect{r},\vect{r}_{0},t]]|}e^{-\frac{i\pi h_{n}}{2}}$$
with $h_{n}$ counting the number of conjugate points along the n-th orbit. 
The probability amplitude is related to Lyapunov exponents:
$$|\mathcal{A}_{n}|\sim exp\left(-\frac{1}{2}\sum_{\lambda_{k}>0}\lambda_{k}t \right)$$
along unstable orbits;
while
$$|\mathcal{A}_{n}|\sim |t|^{-\nu/2}$$
along stable ones.

The level density of bounded quantum states 
is described by the trace of the propagator.
In semiclassical limit, 
the trace over the propagator
is peaked on 
around the periodic orbits 
and stationary saddle points. 
This allows to semiclassically quantize 
semiclassical unstable periodic orbits
that are densely sited in the invariant set.

In our chaotic system, we expect
many resonances. 
So
that,
transitions' probabilities can be averaged 
over the large number 
resonances' peaks.  
So that, a wavepacket $\psi_{t}(\vect{r})$ 
in a region $R$  ($\nu$-dimensional space)
has a quantum survival probability 
$$P(t)=\int_{R}|\psi_{t}({\bf r})|^{2}d\vect{r}$$
This can be rexpressed in terms of the initial density 
matrix $\rho_{0}=|\psi_{0}\rangle \langle \psi_{0}|$
as 
$$P(t)=tr \mathcal{I}_{D}({\bf r})e^{-\frac{iHt}{\hbar}}\rho_{0}e^{+\frac{iHt}{\hbar}}$$
where $\mathcal{I}_{D}$ is zero for resonances ${\bf r}$ out of the region $D$
and$1$
into $D$. We can 
express the survival probability as
$$P(t)\simeq \int \frac{d\Gamma_{ph}}{(2\pi \hbar)^{f}}\mathcal{I}_{D}e^{{\bf L}_{cl}t}\tilde{\rho}_{0}+O(\hbar^{-\nu+1})$$
$$+\frac{1}{\pi \hbar}\int dE\sum_{p}\sum_{a}\frac{\cos\left(a\frac{S_{p}}{\hbar}-a\frac{\pi}{2}{\bf m}_{p} \right)}{\sqrt{|det({\bf m}_{p}^{a}-{\bf 1})|}}\int_{p}\mathcal{I}_{D}e^{{\bf L}_{cl}t}\tilde{\rho}_{0}dt$$
$$ +O(\hbar^{0})$$
where $d\Gamma_{ph}=d\vect{p}d\vect{r}$,
  the sum is on all the periodic orbits (primary periodic orbits $p$ and the
number of their repetitions $a$);
$S_{p}(E)=\int \vect{p}\cdot d\vect{r}$, $\tau_{p}=\int_{E}S_{p}(E)$,
${\bf m}_{p}$ is the Maslov index, 
and $\mathcal{M}$ is a $(2\nu-2)\times (2\nu-2)$ matrix 
associated to the Poincar\'e map in the 
neighborhood of the a-orbit;
${\bf L}_{cl}$ is the classical Liouvillian operator,
 defined in terms of classical Poisson brackets as 
 ${\bf L}_{cl}=\{H_{cl},...\}_{Poisson}$; $\tilde{\rho}_{0}$
 is the Wigner transform 
 of the initial density state.
 The operator 
 ${\bf L}_{cl}$
have Pollicott-Ruelle resonances as eigenvalues 
$$ {\bf L}_{cl}\phi_{n}=\{H_{cl},\phi_{n} \}_{Poisson}=\lambda_{n}\phi_{n}$$
where eigenstates $\phi_{n}$ are Gelfald-Schwartz distributions.
On the other hand, the adjoint problem 
$$ {\bf L}_{cl}^{\dagger}\tilde{\phi}_{n}=\tilde{\lambda}_{n}\tilde{\phi}_{n}$$
The eigenvalues $\lambda_{n}$  
are in general complex:
their real part $Re(\lambda_{n})\leq 0$
because of they correspond to an ensamble bounded periodic orbits;
their $Im(\lambda_{n})$ correspond to
decays in the ensambles.
Expanding 
the survival probability over resonances as 
$$P(t)\simeq \int \sum_{n}\langle \mathcal{I}_{D}|\phi_{n}(E)\rangle \langle \tilde{\phi}_{n}(E)|e^{\lambda_{n}(E)t}| \phi_{n}(E)\rangle \langle \tilde{\phi}_{n}(E)|\tilde{\rho}_{0}\rangle $$
we can get the 0-th leading order  $\sim e^{\lambda_{0}(E)t}$.
So that, the survival probability 
is behaving like $P(t)\sim e^{-\gamma(E)t}$, {\it i.e} $s_{0}=-\gamma(E)$:
the decay of the system is
 related to the classical escape rate $\gamma(E)$.
  
\vspace{2cm} 

{\large \bf Acknowledgments} 
\vspace{0.1mm}

A.~A would like to thank 
 Gia Dvali, Gerard t'Hooft and Carlo Rovelli  for useful discussions on these aspects. 
A.~A  work was supported in part by the MIUR research
grant "Theoretical Astroparticle Physics" PRIN 2012CPPYP7.

\end{document}